\documentclass[fleqn,usenatbib]{mnras}

\usepackage[T1]{fontenc}
\usepackage{ae,aecompl}
\usepackage{bm}
\usepackage{graphicx}    

\usepackage{amsmath}	
\usepackage{amssymb}	
\usepackage{txfonts}

\title[Forced Eccentricity]{Forced Eccentricity in Circumbinary Discs}

\author[Lubow]{Stephen H. Lubow$^1$\thanks{Email: lubow@stsci.edu}
\\ $^{1}$Space Telescope Science Institute, 3700 San Martin Drive, Baltimore, MD 21218, USA\\}

\date{Accepted XXX. Received YYY; in original form ZZZ}

\pubyear{2022}

\begin{document}
\label{firstpage}
\pagerange{\pageref{firstpage}--\pageref{lastpage}} 
\maketitle

\begin{abstract}
We analyze the eccentric response of a low mass coplanar circumbinary disc to secular tidal forcing 
by a Keplerian eccentric orbit central binary.
The disc acquires a forced eccentricity whose magnitude depends on the properties of the binary and disc.
The  largest eccentricities occur when there is a global apsidal resonance in the disc. The driving frequency by the binary
is its apsidal frequency that is equal to zero. 
A global resonance occurs when the disc properties permit 
the existence of a zero apsidal frequency free eccentric mode. 
 Resonances occur for different free eccentric modes that differ in the number of radial nodes.
For a disc not at resonance,  the eccentricity distribution has somewhat similar form to the eccentricity distributions in discs at resonance that have the closest matching disc aspect ratios.
 For higher  disc aspect ratios, the forced eccentricity distribution in a 2D disc is similar to that of the fundamental free mode.
 The forced eccentricity distribution in a 3D disc is similar to that of higher order free modes, not the fundamental mode, unless the disc is very cool. 
For parameters close to resonance, large phase shifts occur between the disc and binary  eccentricities that are locked
in phase.
Forced eccentricity  may play an important role in the evolution of circumbinary discs and their central binaries.

\end{abstract}


\section{Introduction}
Circumbinary discs play an important role in the evolution of binary
systems. Such discs have been observed around nearby young binaries \citep[see review by ][] {Dutrey2016}.
Circumbinary disc interactions with the central binary have been the subject of many hydrodynamical studies in the contexts of young stars and supermassive black holes
 \citep[e.g.,][]{ Artymowicz1996, Gunther2002, MacFadyen2008, Shi2012, Farris2014, Dorazio2016, Munoz2016, Munoz2019, Heath2020, Dorazio2021,Gutierrez2022}.  
 Such binary-disc interactions can affect the mass and orbital evolution of the binary.
 The properties of such
discs provide important information about the central
binary. 

In this paper, we concentrate on coplanar, prograde binary-disc systems.
Such circumbinary discs have a central cavity or gap that is produced by the tidal
forces of the binary \citep{Artymowicz1994}.
Gas streams flow from the circumbinary disc through the central gap onto  the central binary.
These gas streams are highly time-variable and result in pulsed accretion through the gap.
Such flows replenish the discs that surround each binary member.

Simulations have generally found that the circumbinary discs become eccentric.
In the context of young stars, the disc eccentricity  affects  the dynamics of planetesimal collisions and the
 process of planet formation  within
a disc \citep{Marzari2013, Silsbee2015, Silsbee2021}.
For circular orbit 
binaries,
simulations indicate that the gas streams are the cause of the disc eccentricity, as first suggested by \cite{MacFadyen2008}. Gas streams
get pulled into the gap by the gravitational forces of the binary. But not all the gas that gets pulled
into the gap immediately accretes onto the binary. Towards the end of a pulsed accretion episode,
the gas moves away  from the binary. It gets flung outward and impacts the inner edge of the circumbinary disc,
resulting in  an increase in the circumbinary disc eccentricity. Perhaps  the most  convincing
evidence in favor of this  model comes from simulations that have a large enough central
mass sink to allow for gas inflow in the gap, while suppressing the outflow of the gas flung outward
\citep[see figure 17 in][]{Shi2012}.
These simulations show that eccentricity growth ceases once a large sink is introduced that suppresses
the outflow.
The eccentricity is described  as a precessing mode that is trapped between the disc  inner
edge and a radius of a  few times the inner edge radius \citep{Shi2012, Munoz2020}.

The cause of the eccentricity  in circumbinary discs involving eccentric orbit binaries 
is currently not clear. It could be generated by the effects of gas streams or Lindblad resonances
due to the binary \citep{Lubow1991, Teyssandier2016, Miranda2017}.
Another possibility, explored in this paper, is that the disc  eccentricity is due to the secular effects of 
the eccentric orbit binary. The binary produces a secular potential of eccentric
form (azimuthal wavenumber $m=1$ with slow or no time dependence). Provided that the binary orbit is Keplerian,
this potential is static in the inertial frame and therefore has an apsidal precession  frequency value of zero.
In the absence of a disc, a circumbinary test particle experiences a forced eccentricity due to this potential,
as has been explored in the context of circumbinary planetesimals and planets \cite[e.g.,][]{Moriwaki2004,  Leung2013}.
For small binary eccentricity, the magnitude  of the forced eccentricity is linearly proportional to the binary
eccentricity and inversely proportional to apsidal precession rate of the test particle due to the
binary. The forced eccentricity of a test particle does not precess relative the direction of the binary eccentricity.

In this paper we explore how a disc with gas pressure responds to such forcing by a central eccentric orbit binary.
A related issue is how such a disc responds to an apsidal resonance, as has been 
considered in the past \citep{Ward1998}. At such a local resonance, the apsidal precession rate of the central binary matches the precession rate of a fluid element in the disc at some radius. We show that instead of local apsidal resonances,  the disc is subject to global apsidal
resonances whose properties are quite different from the previously studied mean motion resonances in a disc
\citep{Goldreich1979}.
The eccentric response of the disc can be understood by the 
effects of these global resonances. The phase of the  disc forced eccentricity
is locked relative to the phase of the binary eccentricity. But the phase difference can be large for discs that are in or sufficiently close to resonance.

The forced eccentricity of a circumstellar disc has been analyzed in the contexts of planet formation in binaries and Be/X-ray binaries by \cite{Paardekooper2008} and \cite{ Okazaki2002}, respectively. These studies adopted similar secular equations to those in this paper but did not explore the role of resonances,

In Section \ref{sec:model} we describe the model we apply and the equations that  we solve.
Section \ref{sec:res} contains the results.
The conditions required for a global resonance are discussed in Section \ref{sec:exist}. Section of \ref{sec:disc} provides a discussion
and Section  \ref{sec:summary} provides the summary.


\section{Model} \label{sec:model}

 We extend the 2D disc eccentricity evolution equation in \cite{Goodchild2006}  
to include secular forcing by an eccentric orbit binary.  The 2D model is useful because some
simulations are carried out in 2D for example with the AREPO \cite[e.g.,][]{Munoz2019}  and DISCO \cite[e.g.,][]{Dorazio2021} codes.
The disc mass is assumed to be sufficiently
small that the binary orbit is  Keplerian and so the binary precession rate is  zero.
We consider a polar coordinate system $(r, \phi)$ centered on the binary center of mass with associated disc velocity
$(u(r,\phi, t),v(r,\phi, t))$, density $\Sigma(r, \phi, t)$, and pressure $P(r,\phi, t)$. The binary consists of two components with masses $M_1$ and $M_2$
and $M_{\rm b} = M_1+ M_2$ and semi-major axis $a_{\rm b}$. We denote binary mass fraction as $\mu = M_{\rm 2}/M_{\rm b}$   and binary eccentricity as $e_{\rm b}$.
The time-independent axisymmetric component of the binary potential that gives rise to  gravitational apsidal precession  of the disc
is $\Phi_{\rm a}(r)$.
Expanding $\Phi_{\rm a}(r)$  for $r$ large and $e_{\rm b}$ small, we have to lowest order that
\begin{equation}
\Phi_{\rm a}(r) = - \frac{G M_{\rm b} \mu (1-\mu) a_{\rm b}^2}{4 r^3} .
\label{phia}
\end{equation}
The eccentric component of the potential $\Phi_{\rm e}(r)$
is defined to have azimuthal wavenumber $m=1$ and be time-independent.
For $r$ large and $e_{\rm b}$ small, we have to lowest order that
\begin{equation}
\Phi_{\rm e}(r)  = \frac{15  e_{\rm b}  G M_{\rm b} \mu (1-\mu) (1-2\mu) a_{\rm b}^3}{16 r^4}.
 \label{phie}
 \end{equation}

The disc eccentricity evolution equation in \cite{Goodchild2006}  describes the evolution of a complex eccentricity denoted by $E(r,t)$ that encodes the components of the eccentricity vector. The real part of $E$ is along the eccentricity vector of the binary, while the
imaginary part is in the perpendicular direction. The magnitude of the disc eccentricity is given by $e=|E|$. The  2D disc has an assumed unperturbed
axisymmetric state involving gas pressure  $P(r)$ and  density  $\Sigma(r)$. The disc is subject to adiabatic perturbations caused by the eccentric motions
that are characterized by adiabatic index $\gamma$.
$E$ is defined such that the velocity perturbations associated with the eccentricity are given by
\begin{eqnarray}
u(r, \phi, t) &=& i r \Omega E(r,t) \exp{(-i \phi)}, \\
v(r, \phi, t) &=& \frac{r \Omega}{2} E(r, t) \exp{(-i \phi)}.
\end{eqnarray}
with Keplerian angular velocity $\Omega(r)$  that  is due to a central point of mass $M_{\rm b}$.
We extend the eccentricity evolution equation to include effects of the potential $\Phi_{\rm e}$   to obtain
\begin{eqnarray}
2 r \Omega \partial_t E &=&    - \frac{i E}{r} \partial_r(r^2 \partial_r \Phi_{\rm a}) + \frac{i E}{\Sigma} \partial_r P   \nonumber
 \\ &+ & 
   \frac{i }{r^2 \Sigma} \partial_r(\gamma P r^3 \partial_r E) + i   \left(\partial_r  \Phi_{\rm e} +   \frac{2 \Phi_{\rm e}}{r} \right). \label{Eevol}  
\end{eqnarray} 
The first term on the RHS is associated with gravitational precession, the second term involves precession due to pressure,
the next  term involves precession by pressure and wave propagation due to pressure, and
the final term is due to secular forcing by the eccentric orbit binary.

We note that if there is no pressure $P=0$ and  if $\partial_t E=0$, then the first and last terms on the RHS of equation (\ref{Eevol}) balance. Applying equations (\ref{phia})
and (\ref{phie}), we have to lowest order in $r$ large and $e_{\rm b}$ small, there is a forced eccentricity of a test particle
\begin{equation}
E_{\rm  p} (r)= \frac{5 a_{\rm b} e_{\rm b} (1-2 \mu)}{4 r}. \label{Ep}
\end{equation}
 Equations (\ref{phia}),  (\ref{phie}), and  (\ref{Ep}) were previously derived in  \cite{Moriwaki2004} and \cite{Leung2013}.
 Equation (\ref{Eevol})  generalizes the analysis of forced eccentricity to include the effects of pressure
in a disc.

To describe a 3D disc, we apply equation (2) of \cite{Teyssandier2016} of similar form to obtain
\begin{eqnarray}
2 r \Omega \partial_t E &=&    - \frac{i E}{r} \partial_r(r^2 \partial_r \Phi_{\rm a})  
+ \frac{i E}{\Sigma} \left(4 - \frac{3}{\gamma} \right) \partial_r P  \nonumber  
  + \frac{3 i E}{r \Sigma} \left(1 + \frac{1}{\gamma} \right)  P    \\ & + &  
                          \frac{i }{r^2 \Sigma} \partial_r  \left[ \left(2-\frac{1}{\gamma}\right) P r^3 \partial_r E \right]  
+ i   \left(\partial_r  \Phi_{\rm e} +   \frac{2 \Phi_{\rm e}}{r} \right).   \label{Eevol3} 
\end{eqnarray} 
In this equation  $P$ is the vertically integrated pressure of the unperturbed disc.

\section{Results}
\label{sec:res}

We apply equations (\ref{Eevol}) and (\ref{Eevol3}) to determine the response of a disc
to secular  forcing by the eccentric orbit binary in 2D and 3D respectively. 
In the limit of small  $e_{\rm b}$ that we have taken  in equations (\ref{phia}) and  (\ref{phie}),
the results for $E$ scale linearly with $e_{\rm b}$. Consequently, we express results for $E$ that are
normalized by $e_{\rm b}$.
We take the binary to have secondary mass fraction $\mu = 0.4$.
 We consider a  broad disc with inner radius
$r_{\rm  i} = 2  a_{\rm b}$ and outer radius $r_{\rm  o} = 100  a_{\rm b}$. We take the disc
to have $\Sigma \propto 1/\sqrt{r}$ and $P \propto 1/r^{3/2}$  which corresponds to a 
disc aspect ratio $h= \sqrt{P/(\Sigma \Omega^2 r^2)}$ that is independent of radius.
We also impose a small level of dissipation through a bulk viscosity parameter
involving the imaginary part  of $\gamma$
 as $-Im(\gamma) = \alpha_{\rm b}=0.01$. We apply $Re(\gamma)=1$ for the 2D model and  $Re(\gamma)=1.4$ for the 3D model.
 
 We seek the steady state response to binary forcing  in equations (\ref{Eevol}) and (\ref{Eevol3}) and therefore set $\partial_t E=0$. The equations then  become  ordinary differential  equations in $r$.
We apply boundary condition $\partial_r E=0$ at the disc inner and outer radii.  This boundary condition is equivalent to requiring that the Lagrangian density perturbation near the disc edge vanishes.
The resulting equations are solved numerically using NDSolve in Mathematica.

\begin{figure}
\includegraphics[width=8 cm]{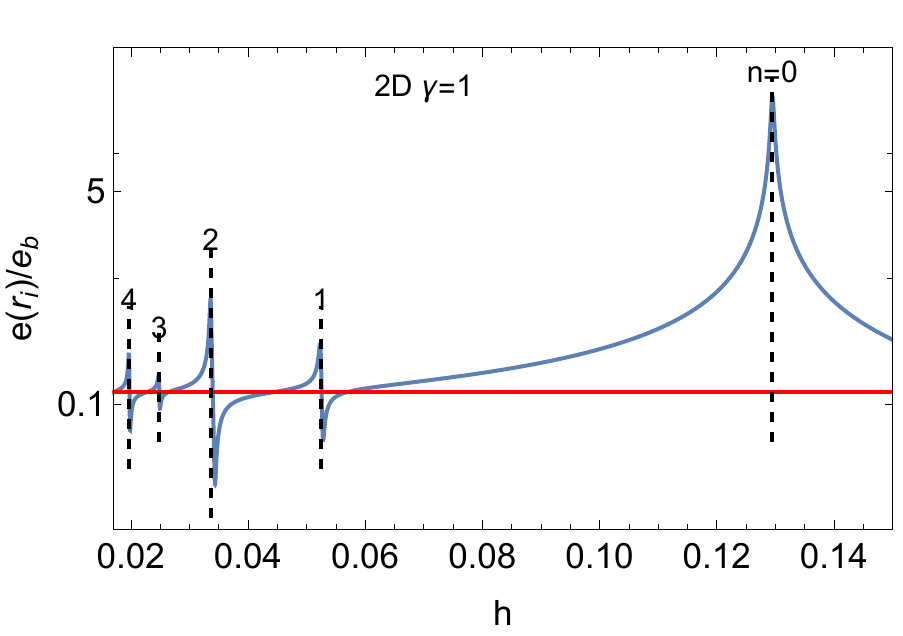}
 \hspace{-0.7cm}
 \includegraphics[width=8 cm]{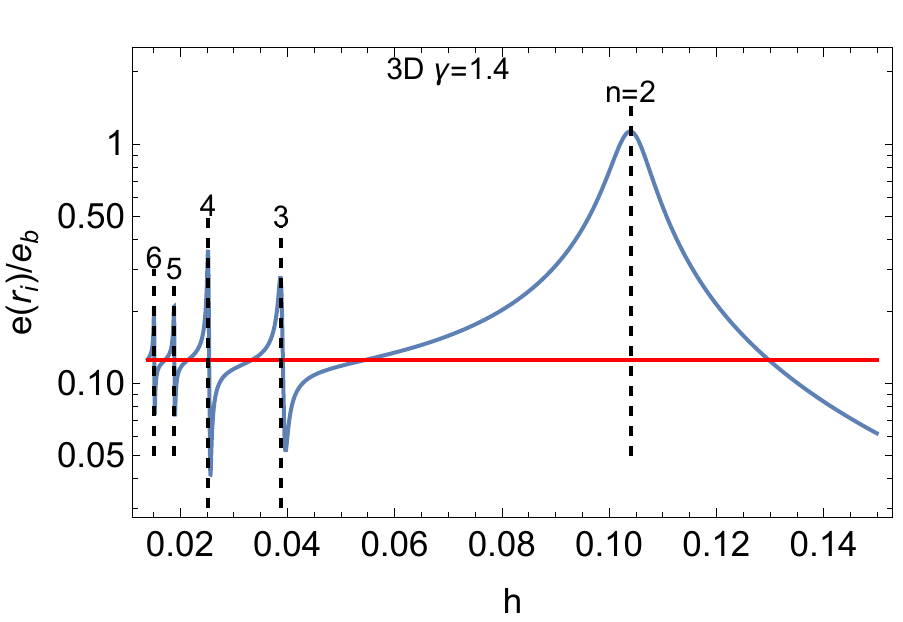}
\caption{The forced  eccentricity at the circumbinary disc inner edge ($r_{\rm i} = 2 a_{\rm b}$)
normalized by the binary  eccentricity  is plotted on a log scale in blue as a function of disc aspect ratio $h$
for the 2D and 3D disc models described in Section \ref{sec:res}. The forced eccentricity
of a test particle at the same radius as  the disc inner edge is plotted on the red line that is determined by  $E_{\rm p}(r_{\rm i})$
in  equation (\ref{Ep}). Each black vertical dashed line occurs at an $h$ value for a free mode at zero frequency where a global resonance
 occurs. The number $n$ with each vertical line is the number of nodes in $E(r)$ for the free mode. } 
\label{fig:eh}
\end{figure}

Figure \ref{fig:eh} plots the magnitude of the forced eccentricity in blue at the disc  inner edge as  a function of
disc aspect ratio $h$ for the 2D and 3D disc models.  The magnitude of the forced eccentricity of a test particle   at the  radius of the
disc inner edge is $e(r_{\rm i})=0.125 e_{\rm b}$  at this same radius  and is plotted
in red.  Notice that the disc forced eccentricity varies considerably and sometimes  achieves values that are much
larger than the test particle value. The 2D and 3D results are qualitatively similar. Both show the strongest
peaks in eccentricity at higher disc  aspect ratios $h \sim 0.1$. The peaks alternate in height for decreasing 
$h$ with the second smaller than the first, while the third is larger than the second, etc.

\begin{figure}
\includegraphics[width=8 cm]{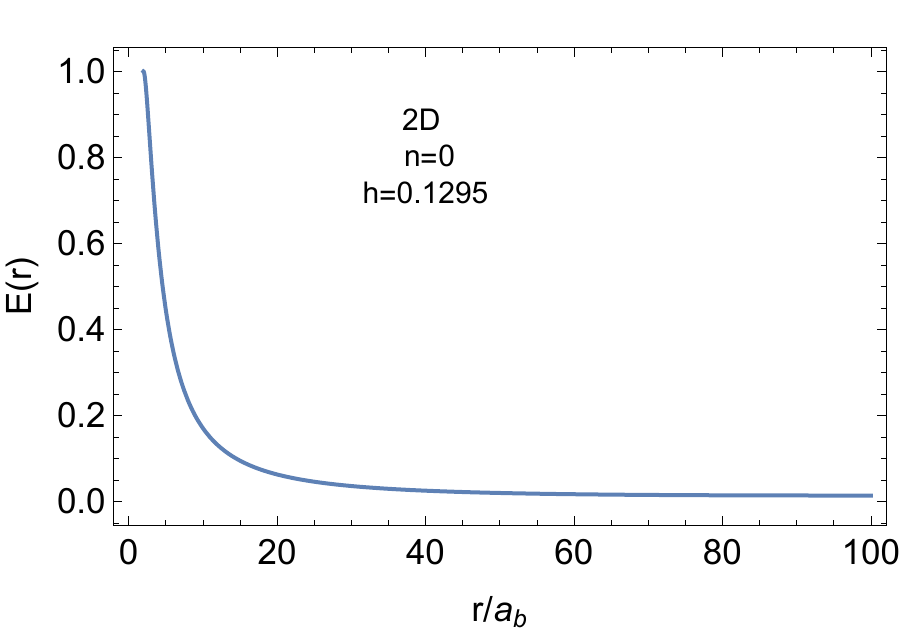}
 \hspace{-0.7cm}
 \includegraphics[width=8 cm]{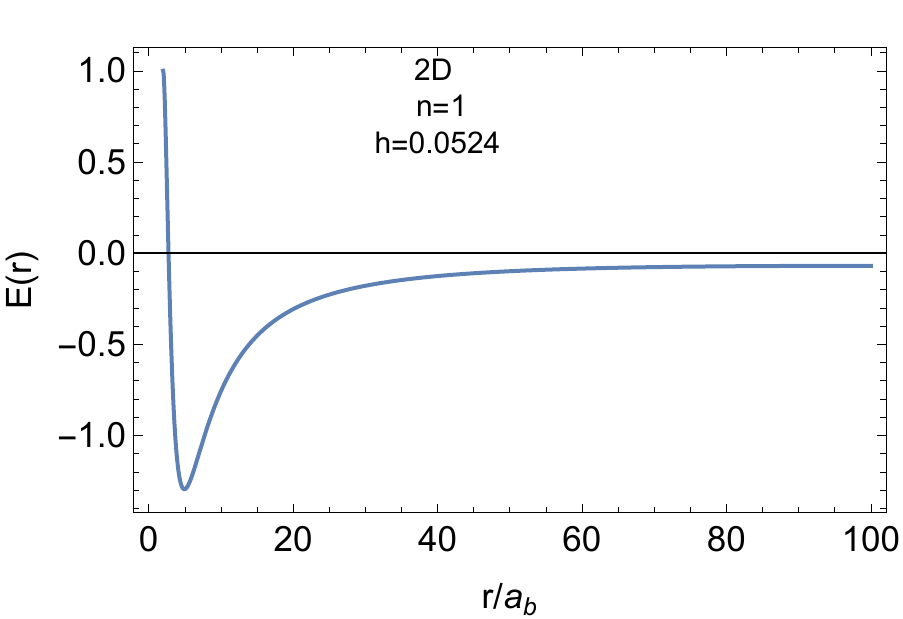}
 \hspace{-0.7cm}
 \includegraphics[width=8 cm]{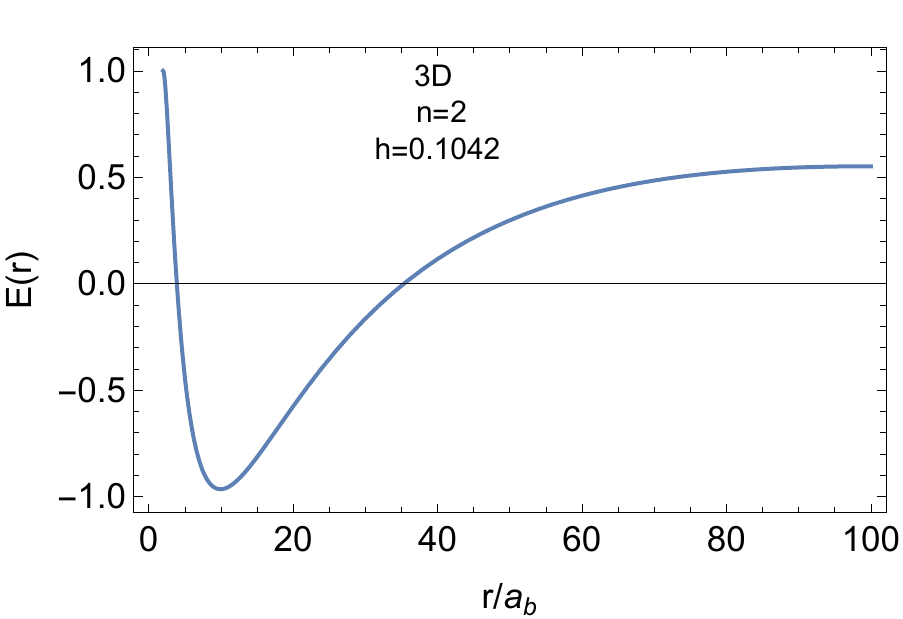}
  \hspace{-0.7cm}
 \includegraphics[width=8 cm]{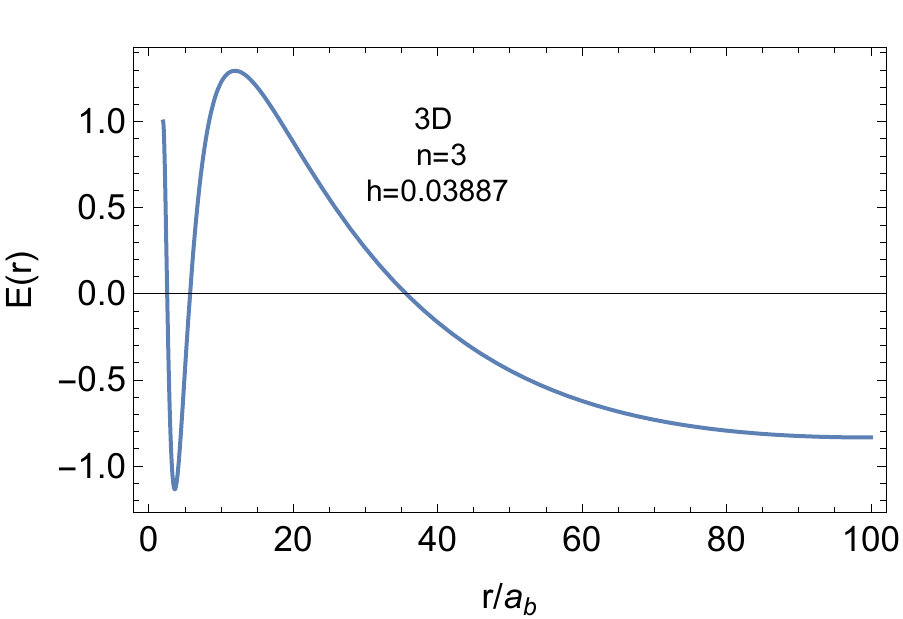}
\caption{The eccentricity distribution for stationary (nonprecessing) free modes that  occur for particular 
disc aspect ratios  $h$
in the 2D and 3D disc models described in Section \ref{sec:res}. The value of $n$  is the number of nodes. These modes are subject to a global resonance due to binary forcing. \label{fig:free}}
\end{figure}

We now demonstrate that the peak eccentricities correspond to the values of $h$ for which a free eccentric mode
has zero precession frequency. The driving of the eccentricity by the binary occurs at its 
apsidal precession frequency of zero. Therefore, the disc undergoes a global apsidal resonance 
 at $h$ values (all other parameters being fixed)
for which the free precession frequency  equals the zero value of the driving frequency.
To compute the free modes, we again solve equations (\ref{Eevol}) and (\ref{Eevol3}) with $\partial_t E=0$ but with  $\Phi_{\rm e}=0$  and without dissipation, $\alpha_{\rm b}=0$.  We apply normalization condition $E(r_{\rm i}) =1$, along with 
$\partial_r E=0$ at the inner and outer boundaries. Free modes with arbitrary values of $h$ generally
have nonzero precession rates, so that $\partial_t E \ne 0$. We determine the values of $h$ for which $\partial_t E=0$,
Some resulting zero frequency 2D and 3D free modes are plotted
in Figure~\ref{fig:free},  There are infinitely many such modes that differ in the number of radial nodes $n$.


 The $h$ values for the five lowest $n$ zero frequency free modes in the 2D and 3D models are plotted as dashed vertical lines in Figure~\ref{fig:eh}.
The plot shows that the $h$ values of the zero frequency modes occur at the forced eccentricity peaks in 
Figures \ref{fig:eh}.  Therefore, we see that these peaks are a consequence of a resonance in which the zero frequency driving matches
the zero frequency of a free mode in the disc.  The peak forced eccentricity at a resonance approaches the test particle value with increasing $n$.
We also find that the peak forced eccentricities vary with bulk viscosity as $ \propto 1/\alpha_{\rm b}$, as would be expected for a resonant effect.

\begin{figure}
\includegraphics[width=8 cm]{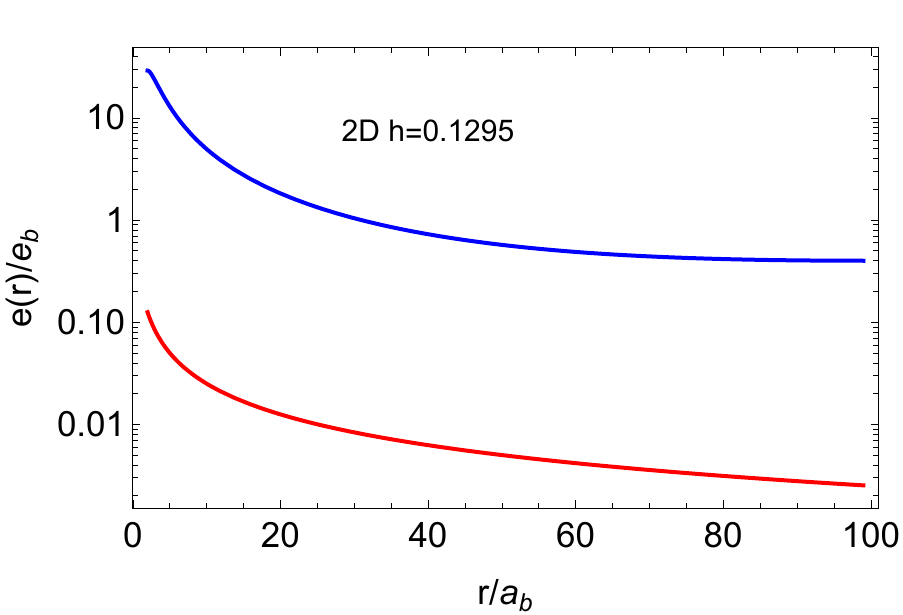}
 \hspace{-0.7cm}
 \includegraphics[width=8 cm]{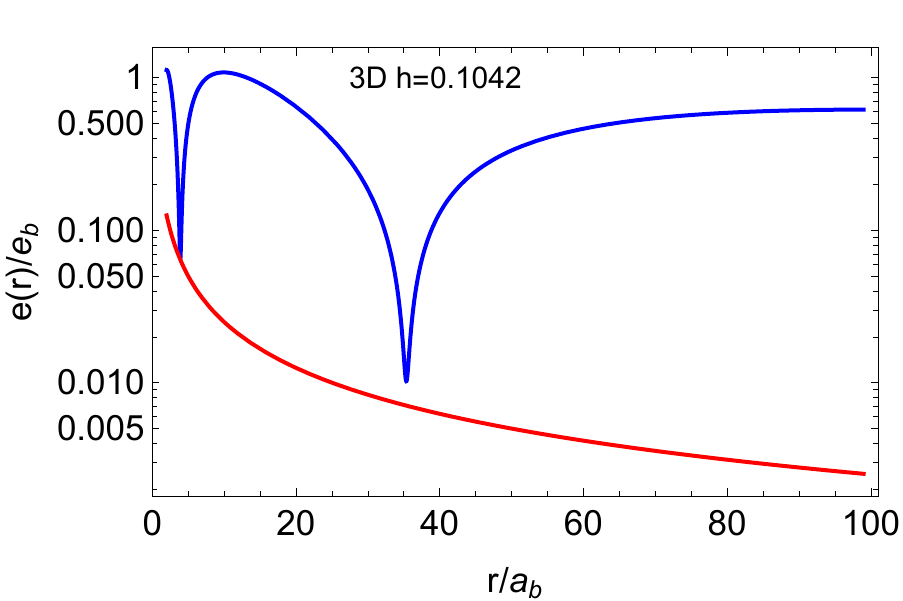}
 \hspace{-0.7cm}
 \includegraphics[width=8 cm]{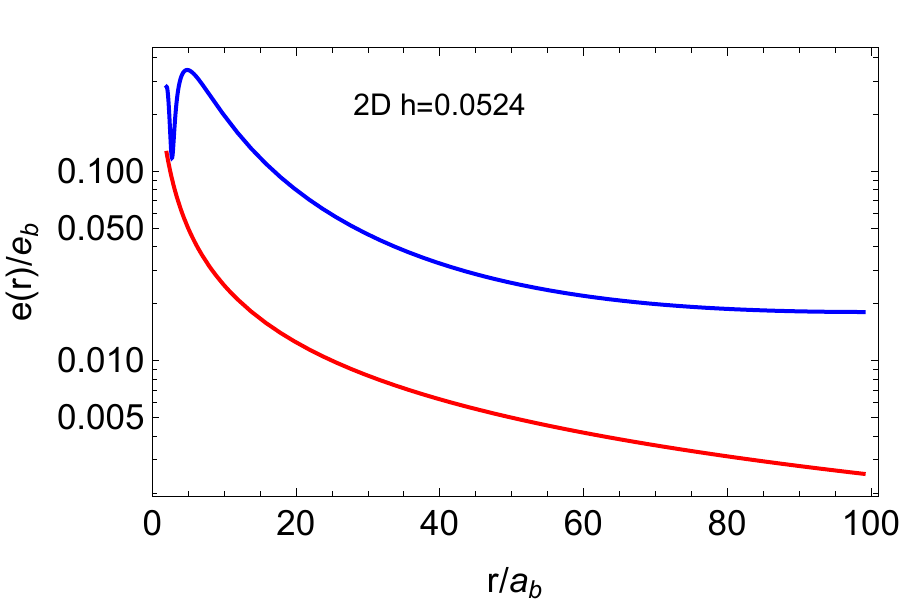}
  \hspace{-0.7cm}
 \includegraphics[width=8 cm]{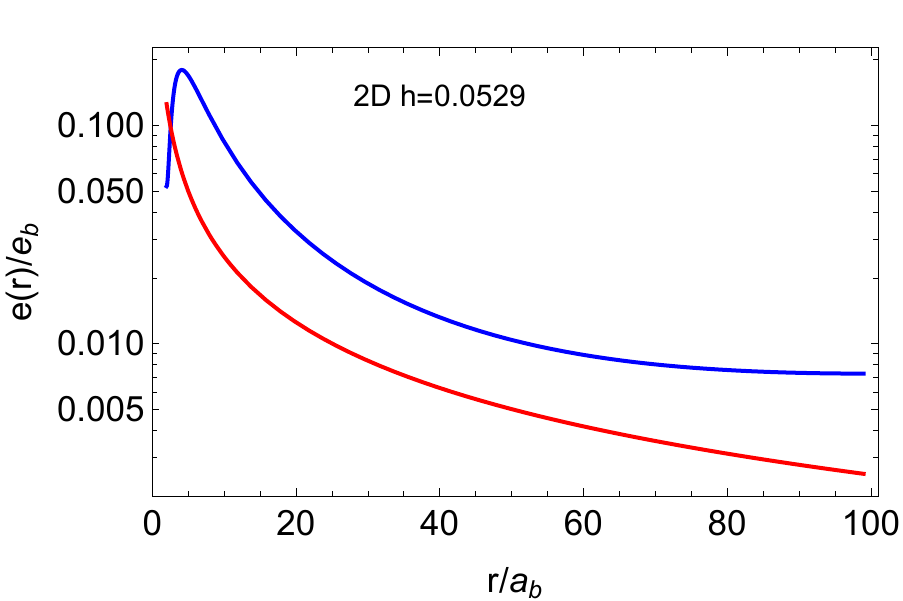}
\caption{The magnitude of the forced eccentricity is plotted on a log scale as a function of radius for the disc (blue) and test
particles (red). The plots are for parameters that are close to resonance. \label{fig:er}}
\end{figure}

Figure \ref{fig:er} plots in blue the  magnitude of the forced eccentricity as  a function of radius
for  the 2D and 3D discs with $h$ values that are close to resonance.  Plotted in red is the
forced eccentricity for test particles. The upper two panels show that shapes of the distributions reflect the distributions
of the resonant free modes that have $n=0$  for the 2D case and $n=2$ for the 3D case, as plotted in 
Figure \ref{fig:free}. The forced eccentricity distributions for the disc generally have much higher values
than for  test particles. For the 2D disc case at $h=0.1295$, the disc 
forced eccentricity distribution
has a similar fall off in radius as  the test particle case. In the 3D  case the disc forced eccentricity distribution is much broader than in the test particle case.  

The bottom two panels of Figure \ref{fig:er} illustrate why there are closely spaced local maximum and minimum values of $e(r_{\rm i})/e_{\rm b}$ near $h=0.0524$ for the 2D model in Figure \ref{fig:eh}. At this $h$ value, there is a zero frequency free mode as seen in  Figure \ref{fig:free}. In the third panel from the top of 
Figure \ref{fig:er}, there are two local maxima for this $h$ value, while at a slightly higher value of $h$
shown in the bottom panel there is a single local maximum and a smaller value of forced eccentricity at the disc inner edge. The transition with decreasing $h$ from single to double maxima then occurs at a zero frequency resonance. This change
is accompanied by a local minimum of $e(r_{\rm i})/e_{\rm b}$ as a function of $h$.
This example explains why there are the pairs of closely spaced local maximum and minimum values
seen in Figure \ref{fig:eh}. Near each resonance there is a change in the number of local maxima in the forced eccentricity distributions.

\begin{figure}
\includegraphics[width=8 cm]{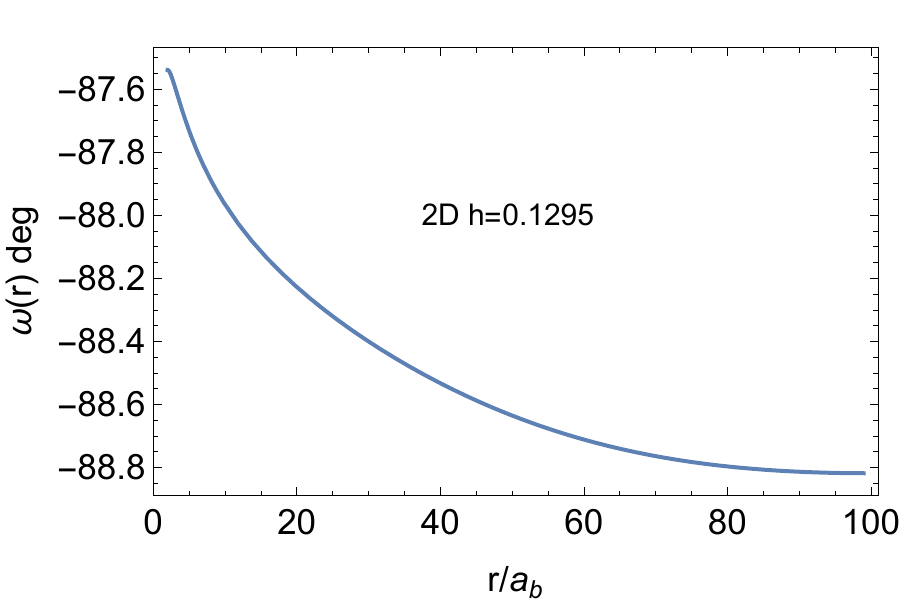}
 \hspace{-0.7cm}
 \includegraphics[width=8 cm]{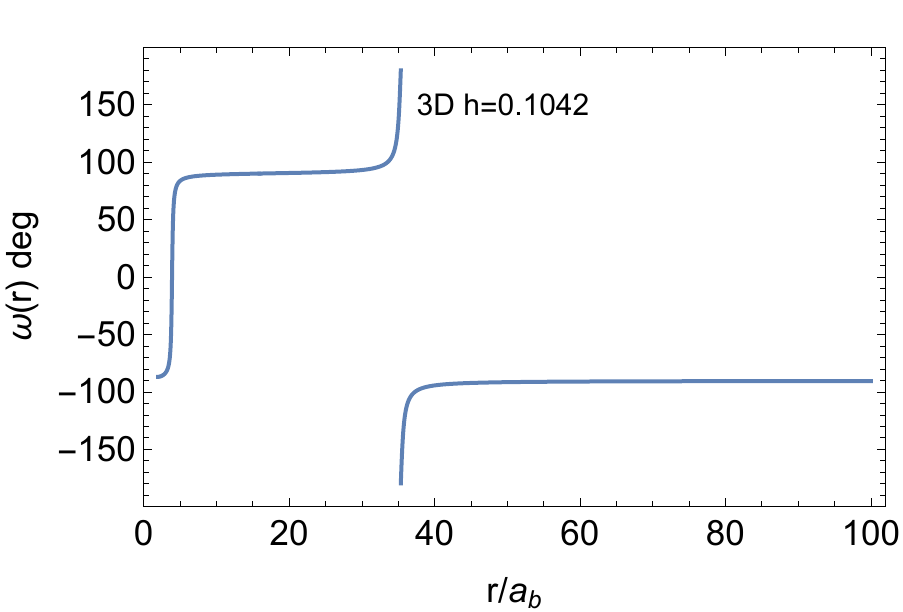}
\caption{The phase of the disc forced eccentricity is plotted as a function of radius for  2D and 3D cases near resonance. \label{fig:phr}}
\end{figure}

Figure \ref{fig:phr} plots the phase of the forced eccentricity as a function of radius for the 2D and 3D cases with parameters
that are close to resonance. In the plotted 2D case, which is for the $n=0$ mode,
 the phase is nearly independent of radius. The phase  varies
considerably in the 3D case that is plotted for the $n=2$ mode. 
The phase changes involves shifts by $180^{\circ}$ at forced eccentricity nodes that are broadened
by viscosity.
Otherwise the phase is fairly constant at $ \simeq \pm 90^{\circ}$.

\begin{figure}
\includegraphics[width=8 cm]{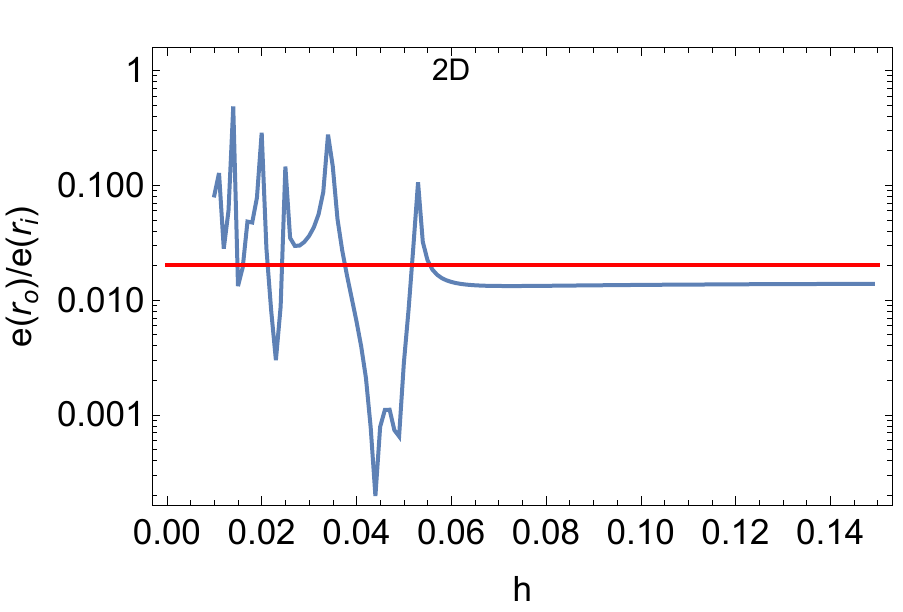}
 \hspace{-0.7cm}
 \includegraphics[width=8 cm]{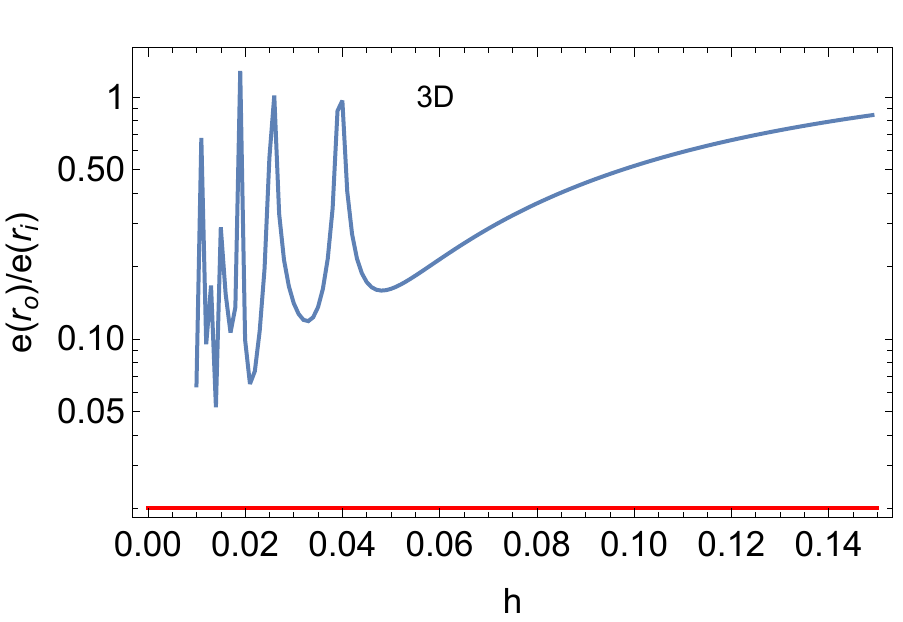}
\caption{The ratio of forced eccentricity at the outer to inner edge of the circumbinary disc  is plotted on a log scale in blue as a function of disc aspect ratio $h$
for the 2D and 3D disc models described in Section \ref{sec:res}. The forced eccentricity ratio 
of  test particles is plotted on the red line  and is equal to 0.02. \label{fig:eroeri}}
\end{figure}

Figure \ref{fig:eroeri} plots in blue the ratio of the forced eccentricity at the outer to inner disc edge as a function of disc aspect ratio $h$ for the 2D and 3D cases. Plotted in red is this ratio for test particles. The plot crudely indicates
the radial extent of the disc forced eccentricity.
 The plot provides another way to show that the distribution of disc forced eccentricity is very different from the test particle eccentricity distribution and is more different in 3D than 2D.
 We have checked that for very small $h<4 \times 10^{-4}$ (not plotted) the disc and test particle ratios are nearly equal, as is expected in the small pressure limit. The structure of the plots is affected by the locations of the disc
resonances evident as peaks in Figure \ref{fig:eh}. For the 2D case with $h \ga 0.06$, the disc eccentricity ratio is consistently smaller than the test particle ratio by about
30\%. While in the 3D disc case with $h \ga 0.06$, the ratio is much larger, indicating that the 3D forced eccentricity distributions are much broader, as is evident
in Figure \ref{fig:er}.  The forced eccentricity ratio is generally larger for the 3D disc case across all disc aspect ratios.

\section{Existence of Zero Frequency Free Modes}
\label{sec:exist}

The properties of the forced eccentricity are determined by resonances in the disc.
Within the framework of the current model, such resonances rely on the existence of zero frequency
free apsidal modes in the disc, as seen in Figure~\ref{fig:eh}. In this section we discuss the conditions
required for such modes to exist.

A useful measure of the effects of disc eccentricity is the angular momentum deficit 
\begin{equation}
A_{\rm d}= \int_{r_{\rm i}}^{r_{\rm o}} \frac {1}{2} |E|^2 \Sigma r^2 \Omega 2 \pi r dr
\label{Ad}
\end{equation}
\citep{Teyssandier2016}. This quantity measures the angular momentum difference of a slightly eccentric disc from a
circular state with the same orbital energy. For the strongest resonance in 2D disc that has  $h=0.1295$ and the 3D disc with $h=0.1042$  in Figure \ref{fig:eh}, the angular momentum deficits
 are $A_{\rm d}= 0.93  e_{\rm b}^2 J_{\rm d}$ and $0.14 e_{\rm b}^2 J_{\rm d}$ respectively, where $J_{\rm d}$
 is the disc angular momentum.

\subsection{2D Modes}

In this subsection we discuss how 2D free modes are able to satisfy the zero frequency condition required for resonance.
By manipulating the equation (\ref{Eevol3}), we calculate the frequency of  free mode in terms of its eigenfunction $E(r)$  
\citep{Goodchild2006}.
The frequency of a free mode can be decomposed as
\begin{equation}
\omega = \omega_{\rm g} + \omega_{\rm p1} + \omega_{\rm p2} ,
\label{omegadecomp}
\end{equation}
where
\begin{eqnarray}
\omega_{\rm g}  &=& -\frac{\pi} {2 A_{\rm d}} \int^{r_{\rm o}}_{r_{\rm i}} \partial_r \left(r^2  \partial_r \Phi_{\rm a} \right)  | E(r)  |^2 \Sigma(r) r  dr, \label{omg2D}\\
 \omega_{\rm p1} &=&   -\frac{\pi \gamma} {2 A_{\rm d}} \int^{r_{\rm o}}_{r_{\rm i}}  | \partial_r E(r)  |^2  P r^3 dr ,  \label{omp12D}\\\
  \omega_{\rm p2} &=&   \frac{\pi} {2 A_{\rm d}} \int^{r_{\rm o}}_{r_{\rm i}}  |E(r)  |^2 \partial_r P   \,  r^2 dr ,  \label{omp22D}\\\
\end{eqnarray}
 $A_{\rm d}$ is the angular momentum deficit defined in equation (\ref{Ad}), $\omega_{\rm g}$ is due to the gravitational
effects of the binary, and $\omega_{\rm p1}$  and  $\omega_{\rm p2}$  are due to pressure effects. In particular,  $\omega_{\rm p1}$ is due to a flux term, the third term on the RHS of equation (\ref{Eevol}). 

Since the gravitational precession term $\omega_{\rm g}$ is positive, the pressure precession terms
must sum to a negative value in order to have a zero frequency mode.  Since both pressure terms are negative, this requirement is satisfied for all $n \ge 0$. In fact, we find that zero frequency free modes exist for all $n$ for suitable values of $h$.

\subsection{3D Modes}
\label{sec:3Dex} 

The existence of zero frequency free modes in 3D is more complicated than in the 2D case.
By manipulating the equation (\ref{Eevol3}), we calculate the frequency of  free mode in terms of its eigenfunction $E(r)$  
\citep{Teyssandier2016}.
The frequency of a free mode can be decomposed as
\begin{equation}
\omega = \omega_{\rm g} + \omega_{\rm p1} + \omega_{\rm p2} +\omega_{\rm p3},
\label{omegadecomp}
\end{equation}
where
\begin{eqnarray}
\omega_{\rm g}  &=& -\frac{\pi} {2 A_{\rm d}} \int^{r_{\rm o}}_{r_{\rm i}} \partial_r \left(r^2  \partial_r \Phi_{\rm a} \right)  | E(r)  |^2 \Sigma(r) r  dr, \\
 \omega_{\rm p1} &=&   -\frac{\pi} {2 A_{\rm d}} \left(2-\frac{1}{\gamma}\right)  \int^{r_{\rm o}}_{r_{\rm i}}   |\partial_r E(r)  |^2  P r^3 dr ,\\
  \omega_{\rm p2} &=&   \frac{\pi} {2 A_{\rm d}}  \left(4 - \frac{3}{\gamma} \right)  \int^{r_{\rm o}}_{r_{\rm i}} |E(r)  |^2 \partial_r P   \,  r^2 dr ,\\
    \omega_{\rm p3} &=&   \frac{3 \pi} {2 A_{\rm d}}  \left(1 +\frac{1}{\gamma} \right) \int^{r_{\rm o}}_{r_{\rm i}}  |E(r)  |^2   P   \,  r dr ,\\
\end{eqnarray}
 $A_{\rm d}$ is the angular momentum deficit defined in equation (\ref{Ad}), $\omega_{\rm g}$ is due to the gravitational
effects of the binary, and $\omega_{\rm pi}$  for $i=1,2,3$ are due to pressure effects. 
In particular,  $\omega_{\rm p1}$ is due
to a flux term, the fourth term on the RHS of equation (\ref{Eevol3}).
Frequencies $\omega_{\rm g}$, $ \omega_{\rm p1}$, and $ \omega_{\rm p2}$ are similar to their 2D counterparts in equations (\ref{omg2D}),
 (\ref{omp12D}), and  (\ref{omp22D}), respectively,
and are
in fact equal for $\gamma=1$. Frequency $ \omega_{\rm p3}$ is due to 3D effects.
 We have numerically verified that equation (\ref{omegadecomp})
is satisfied with $\omega=0$ to machine accuracy for the 3D  zero frequency adiabatic free modes in this paper.

Since the gravitational precession term $\omega_{\rm g}$ is positive, the pressure precession terms
must sum to a negative value in order to have a zero frequency mode. Pressure term is $\omega_{\rm p3}$
 is the only positive pressure contribution. However, it is not always small.
For $P \propto r^{-s}$, it follows that  $\omega_{\rm p2} + \omega_{\rm p3} < 0$ requires that
$s  > 3(1+\gamma)/(-3+4 \gamma)$. This is condition is typically not satisfied, as is the case for
the parameters in this paper that have $s=1.5$ and $\gamma=1.4$. An overall negative pressure induced precession is possible
due to term $\omega_{\rm p1}$. This term becomes more important for higher $n$ modes for which the radial
derivatives of eccentricity become increasingly important.

For the model we have considered,  the fundamental mode $n=0$, as well as $n=1$, have only positive eigenfrequencies for any
reasonable value of $h<0.2$. This result is consistent with the findings of \cite{Miranda2018} who computed
eigenfrequencies (precession frequencies) of the fundamental free mode in a similar configuration and reported only positive values.

\begin{figure}
\includegraphics[width=8 cm]{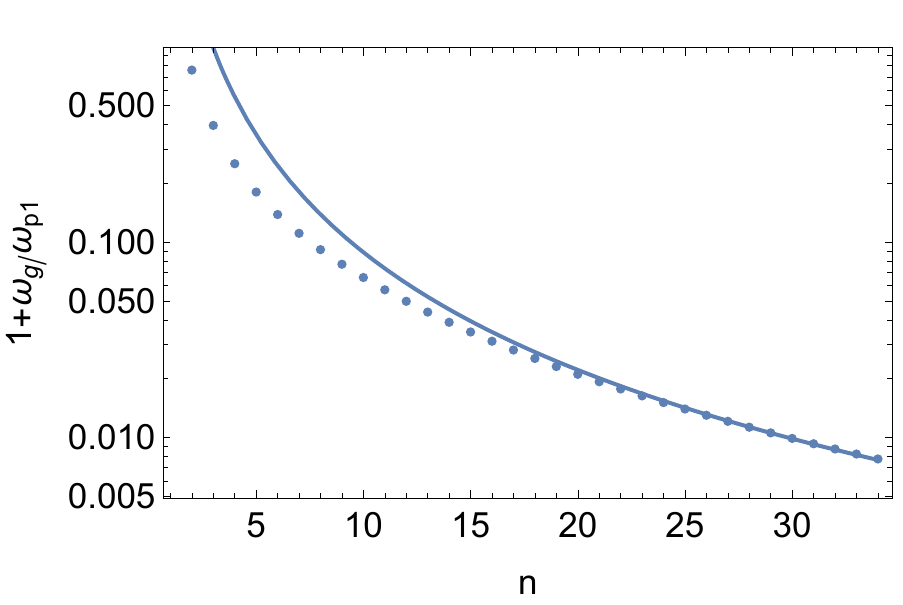}
\caption{ $1+\omega_{\rm g}/ \omega_{\rm p1}$ plotted on a log scale against mode number $n \ge 2$ for zero frequency modes. The points plot the numerically determined values
for each $n$. The disc aspect ratio decreases with increasing $n$. The solid line plots $8.86/n^2$.
\label{fig:omgomp1}}
\end{figure}

With the parameters in this paper, zero frequency modes exist for modes with $n \ge 2$  due to the negative contribution of $\omega_{\rm p1}$.
This term becomes more important with increasing $n$. As a result, lower pressure and thus lower $h$ values are required
to obtain a zero frequency mode. This effect explains why the resonant peaks in Figure~\ref{fig:eh} occur at smaller $h$ with increasing $n$.
 Figure~\ref{fig:omgomp1}  shows that $\omega_{\rm p1}+\omega_{\rm g} \sim 8.86 \omega_{\rm p1}/n^2 < 0 $ for zero frequency modes with large $n$. 
The other pressure contributions, $\omega_{\rm p2}+\omega_{\rm p3}$, become unimportant at large $n$.

\section{Discussion}
\label{sec:disc}

Mean motion resonances between  perturber  and a gas disc can play an important role in disc dynamics and eccentricity generation
\citep{Goldreich1979, Goldreich1980, Lubow1991}.   At such a resonance, the absolute value of the Doppler shifted frequency of a component of the forcing by the perturber matches the epicyclic frequency in the disc.
In particular, the Lindblad resonances launch waves from a region that is of small radial extent  (that scales with $h^{2/3}$)
compared to the  radius at the resonance.
These waves then propagate and damp as they travel away from the resonance. The smallness
of the wave generation region comes about  because  the frequency associated with such a mean motion
resonance  is high compared to frequency associated with the gas sound speed $c_{\rm s}/r$ for a thin disc.
The radial width of the resonance  depends then on the disc aspect ratio which is small.
In the application of this resonance model to the case of secular forcing,  
 the resonance radius  would occur  where the frequency of the secular forcing matches
 the apsidal precession frequency  in the disc \citep{Ward1998}.  Since the strength of the  Lindblad torque depends inversely on the frequency associated with a resonance,
 this model predicts that very strong torques are produced by  the low frequency  apsidal motions.
 
 As pointed out  by \cite{Goldreich2003}, the  behavior of waves involving  an apsidal resonance is quite
 different from the behavior in the mean motion resonance case. Because of their low frequency, apsidal waves are not 
 generated in a small region of space. Instead such waves are better understood as global standing waves.
 Our results support that view. 
 A similar situation occurs in the secular dynamics of tilted discs in binaries.
 Secular nodal resonances are also greatly broadened by gas pressure to the extent that the usual mean motion
torques do not apply \citep{Lubow2001}.
 The  resonances we find here are  global
and occur for particular  disc  parameters, unlike the Lindblad resonances.  
Unlike the mean motion Lindblad resonance, this resonance does not provide a local site for wave launching.
The  disc forced eccentricity is affected by the resonance 
over
the full radial extent of the mode.

\begin{figure}
\includegraphics[width=8 cm]{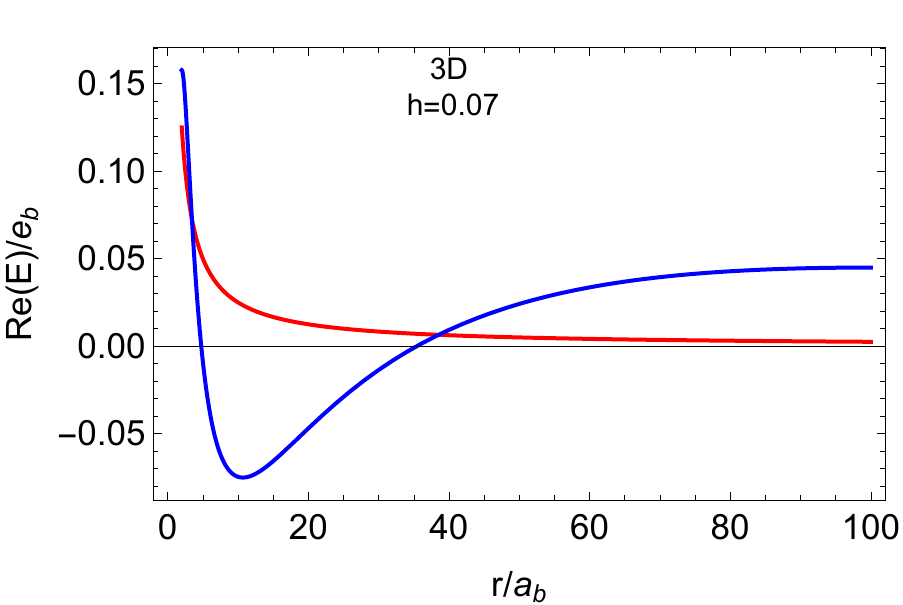}
 \hspace{-0.7cm}
 \includegraphics[width=8 cm]{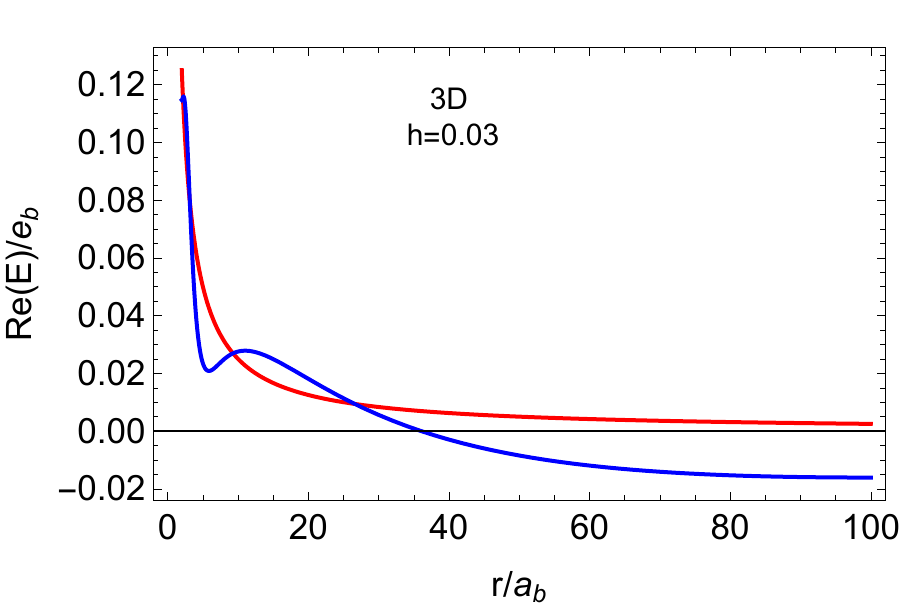}
 \hspace{-0.7cm}
 \includegraphics[width=8 cm]{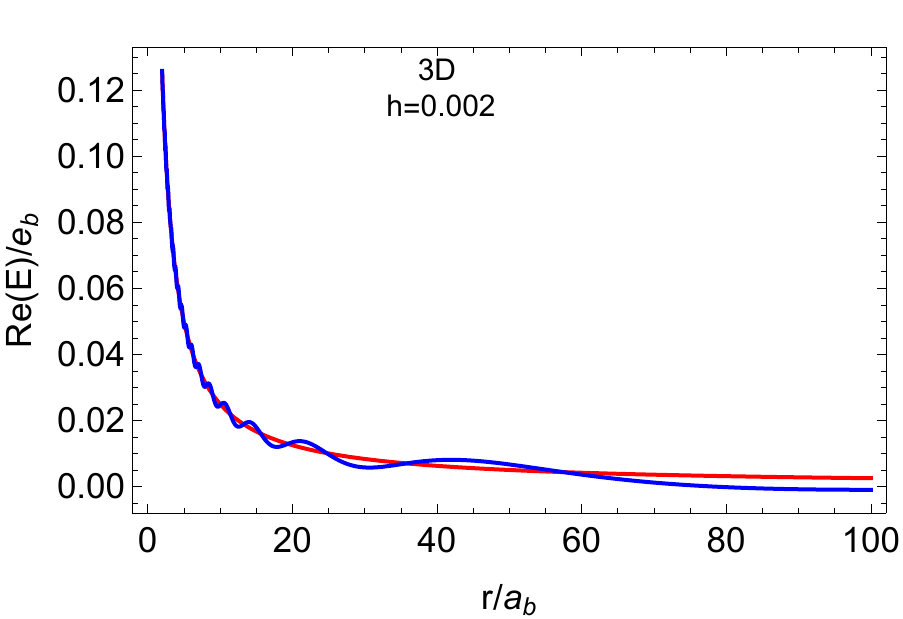}
\caption{The real part of the  forced eccentricity is plotted for the disc (blue) and 
test particles (red). The plots are for $h$ values that are between resonances.
$Im(E)$ is much smaller in absolute value than $|Re(E)|$ and is not plotted.
 \label{fig:enr}}
\end{figure}
 

 The size of the inner cavity depends on binary eccentricity \citep{Artymowicz1994, Pelupessy2013}. If the size of the inner cavity is changed from our assumed value of $r_{\rm i}=2 a_{\rm b}$, the resonances occur at different $h$ values from those in Figure~\ref{fig:eh}.
For example, if  $r_{\rm i}$ is increased to   $3 a_{\rm b}$ , the $n=2$ resonance in the 3D case decreases from $h=0.10$ to 0.05. This is expected since
the gravitational precession is weaker and so lower pressure is required for a zero frequency mode.

The torque associated with the forced eccentricity damping is given by
\begin{equation}
T = -\frac{\alpha_{\rm b}}{2} \int_{r_{\rm o}}^{r_{\rm o}} P r^2 \left|\frac{d E}{dr} \right|^2 2 \pi r dr
\end{equation}
that is equal to the rate of change of angular momentum deficit for free waves given by equation (\ref{Ad}) \citep{Teyssandier2016}.
This torque results in a binary eccentricity decrease and fixed semi-major axis.
For the strongest resonance in 2D disc and the 3D disc plotted in  Figure \ref{fig:eh},  $d ln(e_{\rm b})/dt = -1.4  \times 10^{-5}  \,   \Omega_{\rm b} J_{\rm d}/  J_{\rm b}$ and $-1.7 \times 10^{-7} \,   \Omega_{\rm b} J_{\rm d}/  J_{\rm b}$ respectively. 
for these two cases respectively.  For example, for 
$\alpha_{\rm b}=0.01$ and  $J_{\rm d} = 0.1 J_{\rm b}$, the timescale to change the binary eccentricity is  about $1 \times 10^5$ and $9  \times 10^6$ binary orbital periods for these two cases. The results  suggest that in the 3D case significant changes
in binary eccentricity over a disc lifetime of $\sim 10^6$ yr can occur only if the binary period is fairly short, $< 0.1$ yr. The timescale to change the binary eccentricity at resonance depends linearly on $\alpha_{\rm b}$. More  complete calculations would take into account the effects  of  nonzero binary precession as a consequence of significant disc mass,  as
has been studied in the context of circumstellar discs by \cite{Silsbee2015}.

Simulations have been carried out that explore circumbinary disc eccentricity evolution involving eccentric orbit binaries 
\citep[e.g.,][]{Miranda2017}. Recently, \cite{Siwek2022} analyzed  circumbinary disc  eccentricity evolution using 2D simulations that covered a set of
binary mass ratios and eccentricities. 
The results show disc eccentricity phase locking at large angles relative to the binary eccentricity, as we 
find here near resonances.  On the other hand,   phase locking of disc  eccentricity 
 was found in one case involving an equal mass  eccentric orbit binary. Forced disc eccentricity does not occur for an equal mass binary 
 that has $\mu=1/2$ (see equation (\ref{Ep})). 

The forced eccentricity distributions that result for $h$ values between resonances in Figure~\ref{fig:eh} 
are strongly affected by the structure of the closest resonant modes, even when the disc forced eccentricity is not significantly  amplified  by a resonance.  In the case of the 2D disc, the forced eccentricity distribution for $h> 0.06$ resembles the fundamental mode  (see Figure~\ref{fig:free}) due to
 the presence of the $n=0$ resonance. For smaller values of $h$ there are more complicated structures in the forced eccentricity distribution, due to effects of weak higher order resonances. For very small values of $h < 0.005$, the disc forced eccentricity distribution approaches that of test particles.

 The nonexistence of a resonance involving the fundamental mode in 3D, as discussed in Section \ref{sec:3Dex}, has interesting consequences.
Consider for example $h=0.07$ that lies between the $h$ values of the $n=2$ and $n=3$ resonances shown in  the bottom panel of Figure~\ref{fig:eh}. In the top panel of Figure~\ref{fig:enr},
we see that the forced eccentricity distribution resembles the $n=2$ free mode that is associated with the nearby resonance at $h=0.1042$ (see Figure~\ref{fig:free}), not the fundamental mode.  
This property holds for more generally for $h>0.045$.
The value $h=0.03$ lies between the $n=3$ and $n=4$ resonances in the bottom panel of Figure~\ref{fig:eh}. In the middle panel of Figure~\ref{fig:enr},
we see that the forced eccentricity distribution  is  again not like the fundamental. 
For a much cooler disc with $h=0.002$ that lies
between $h$ values  for the $n=41$ and $n=42$ resonances, the forced eccentricity 
 distribution is plotted in the bottom panel of Figure~\ref{fig:enr}. There are wiggles resulting from the weak  high $n$ resonances. The
 overall forced eccentricity  distribution is similar that of  test particles. For even cooler discs, the forced eccentricity distributions are indistinguishable
 from the  test particle forced eccentricity distribution. Such cool discs may be relevant for the case involving supermassive black hole binaries.
 
 On the other hand, the eccentricity distribution in an unforced 3D disc with some arbitrary initial eccentricity
distribution 
would be expected to evolve over time to the fundamental mode distribution because 
higher $n$ modes are expected to decay faster  \citep[][]{Miranda2018}. However, we find that the situation is quite
different for forced eccentricity in 3D, in which the fundamental mode distribution  occurs only for very cool discs, $h < 0.01$. More generally, which eccentric disc mode is excited also depends on the properties of the forcing. In the case of superhump binaries in which the circumstellar disc eccentricity is excited at the 3:1 resonance, the fastest growing disc mode depends on the disc outer radius and is  not necessarily the fundamental \citep[see Fig. 8 of][]{Lubow2010}.

\section{Summary}
\label{sec:summary}

We have analyzed the response of a gaseous circumbinary disc to secular forcing by an eccentric orbit
central binary through the application of the disc eccentricity equations of \cite{Goodchild2006} and \cite{Teyssandier2016}.  
The disc acquires a forced eccentricity that can significantly exceed the corresponding values
for  test particles (see Figures  \ref{fig:eh} and \ref{fig:er}). The disc forced eccentricity is enhanced due to global apsidal resonances.
 The strongest enhancements occur for higher values of disc aspect ratios.
At these resonances, the precession rate of a free disc mode matches the precession rate of the central
binary that is zero in the case of a low mass  disc. Due to the width of these 
resonances, the forced eccentricity is also enhanced for conditions near those for a resonance. Large phase
shifts occur between the forced eccentricity of the disc and binary near resonance  (Figure \ref{fig:phr}). The phase shifts sometimes
vary in radius. The forced eccentricity distributions in 3D are typically broader than in the 2D case  (Figure \ref{fig:eroeri}).
 Near each resonance there is a change in the number of local maxima in the forced eccentricity distributions (Figure \ref{fig:er}).

For a disc not at resonance,  the eccentricity distribution has somewhat similar form to the eccentricity distributions in discs at resonance that have the closest matching disc aspect ratios. For a 2D disc, the forced eccentricity distribution is similar to that of the fundamental mode over a broad range of parameters at higher disc aspect ratios $h > 0.05$.
 In the case of a 3D disc, the forced eccentricity distribution is similar to that of higher order free modes, not the fundamental, unless the disc is very cool ($h < 0.01$), as seen in Figure~\ref{fig:enr}. This result is a consequence of  the nonexistence of a zero frequency fundamental free mode in a 3D disc. For very cool discs, which may be of relevance to discs around supermassive black hole binaries, the nonresonant forced eccentricity distribution is similar to that of test particles, due to the low pressure. 
 
The conditions required for resonance will change with parameter changes such as binary mass ratio and eccentricity,
as well as the disc pressure variation in radius. 
At higher order, there is a dependence
of  potential $\Phi_{\rm a}$ in equation (\ref{phia}) on binary eccentricity that affects  the conditions for resonance.
In addition,  the forcing potential 
$\Phi_{\rm e}$ in equation (\ref{phie})  varies with binary eccentricity beyond the lowest order linear 
approximation that we have adopted. 
For a disc of significant mass, the binary will undergo precession which in turn will modify the
conditions required for resonance.
Self-gravity could also modify the disc precession rate.  In addition, the binary orbit might
be modified by its  secular interaction  with the circumbinary disc. 
This study did not include the effects of shear viscosity that could
play a role in causing additional phase shifts between the disc and binary. If high disc eccentricities are achieved, nonlinear dissipation though shocks may occur \citep[e.g.,][]{Martin2014}.
While this paper has been concerned with circumbinary discs,
similar resonance effects might occur with the secular eccentric forcing of circumstellar discs.

\section*{Acknowledgements}
 I thank the referee for carefully reading the paper and raising insightful questions.
I thank Eugene Chiang, Diego  Mu\~{n}oz, Gordon Ogilvie, and  Magdalena Siwek for useful discussions.
 Most of this work was carried out at the KITP  Binary22 program.
 I acknowledge support from NASA  grants 80NSSC21K0395
and 80NSSC19K0443. This research was supported in part by the National Science Foundation under Grant No. NSF PHY-1748958.

\section*{Data availability}
The data underlying this article will be shared on reasonable request to the author.

\appendix



\bibliographystyle{mnras}
\bibliography{ref} 

\bsp	
\label{lastpage}
\end{document}